\documentclass[10pt,twocolumn]{article}
\usepackage{authblk}
\usepackage{graphicx}
\usepackage{amsmath,amssymb}

\begin{document}

\title{The band gap problem: the accuracy of the Wien2k code confronted.}
\author{J. A. Camargo-Mart\'inez and R. Baquero}
\affil{{\it Departamento de F\'isica}, CINVESTAV-IPN, Av. IPN 2508, 07360 M\'exico}

\maketitle

\begin{abstract}
This paper is a continuation of our detailed study [Phys. Rev. B 86, 195106 (2012)] of the performance of the recently proposed modified Becke-Jonhson 
potential (mBJLDA) within the known Wien2k code. From the 41 semiconductors that we have considered in our previous paper to compute the band gap value,
we selected 27 for which we found low temperature experimental data in order to pinpoint the relative situation of the newly proposed Wien2k(mBJLDA) method
as compared to other methods in the literature. We found that the GWA gives the most accurate predictions. The Wien2k (mBJLDA) code 
is slightly less precise, in general. The Hybrid functionals are less accurate, on the overall. The GWA is definitely the most precise existing method 
nowadays. In 88\% of the semiconductors considered the error was less than 10\%. Both, the GWA and the mBJLDA potential, reproduce the band gap of 15 
of the 27 semiconductors considered with a 5\% error or less. An extra factor to be taken into account is the computational cost. If one would seek for 
precision without taking this factor into account, the GWA is the method to use. If one would prefer to sacrifice a little the precision obtained against 
the savings in computational cost, the empirical mBJLDA potential seems to be the  appropriate method. We include a graph that compares directly the performance 
of the best three methods, according to our analysis, for each of the 27 semiconductors studied. The situation is encouraging but the problem is not yet
a closed issue.

\bigskip
{\it Keywords}: Band gap problem; Wien2k; mBJLDA potential; hybrid functionals; GW approximation.

\smallskip
PACS: 71.15.Mb; 71 .20.Mq; 71. 20.Nr
\end{abstract}



\section{Introduction}

Khon-Sham equations~\cite{1} are central to the practical application of Density Functional Theory (DFT). To solve them, an approximation to the exchange 
and correlation energy is required from which an exchange and correlation potential is derived. The way in which this term is approximated is crucial to the 
proper description of the band structure of solids. The Local Density Approximation (LDA)~\cite{2}, the Generalized Gradient Approximation (GGA)~\cite{3,4,5} 
and the meta-GGA~\cite{3,6}, among others, describe very well  the electronic band structure of even complicated metallic systems. They fail, nevertheless to 
account for the band gap value of semiconducting systems, a short come known for  several years now~\cite{7}.
Efforts to solve this problem were done since long ago. Approximations as the ``scissor operator"~\cite{8}, the Local Spin Density Approximation, 
LSDA+U~\cite{9} and methods based on the use of Green's functions and perturbation theory as the GW approximation, GWA~\cite{10,11,12}, were proposed. 
In the last ten years, these efforts gave rise to substantially improved results.
Some of the new proposals include, the screened hybrid functional of Heyd, Scuseria and Ernzerhof (HSE)~\cite{13,14,15} and the middle-range exchange and correlation 
hybrid functional of Henderson, Izmaylov, Scuseria and Savin (HISS)~\cite{16,17}. Another recent proposal is the empirical potential the modified 
Becke-Jonhson potential (mBJLDA) proposed by Tran and Blaha~\cite{18}. This potential was introduced to the Wien2k code~\cite{19} in 2010.

\section{The mBJLDA potential}

Recently, we made a detailed analysis of the mBJLDA potential based on the calculation of the electronic band structure of 41 semiconductors~\cite{20}. This paper
is a continuation of that work. We found an important improvement in the predictions of the band gap as compared to experiment. The mBJLDA potential~\cite{18} is a 
empirical potential of the form
\begin{equation}
\label{eq:eq2}
V_{x,\sigma}^{MBJ}(r)=cV_{x,\sigma}^{BR}(r)+(3c-2)\frac{1}{\pi}\frac{\sqrt{5}}{12}\frac{\sqrt{2t_{\sigma}(r)}}{\rho_\sigma(r)}
\end{equation}
where $\rho_\sigma(r)$ is the spin dependent density of states, $t_\sigma(r)$ is the kinetic energy density of the particles with spin $\sigma$. and 
$V_{x,\sigma}^{BJ}(r)$ is the Becke-Roussel potential (\emph{BR})~\cite{21}. The c stands for
\begin{equation}
\label{eq:eq3}
c=\alpha + \left(\beta\frac{1}{V_{cell}}\int{d^3 r\frac{\mid \nabla \rho(r)\mid}{\rho(r)}}\right)^{1/2}
\end{equation}
$\alpha$ and $\beta$ are free parameters. The Wien2k code defines  $\alpha=-0.012$ and $\beta=1.023$ $Bohr^{1/2}$. These values are general but 
certainly fixed experimenting with several cases.
A particular feature of this potential is that a corresponding exchange and correlation energy term , $E_{xc}[\rho]$, such that the mBJLDA potential is obtained 
in the usual way, namely, $V_{xc}=\delta E_{xc}[\rho]/\delta \rho$, is not possible. As a consequence, a consistent optimization procedure to obtain the lattice 
parameters, the Bulk modulus and its derivative with respect to pressure are not actually possible. This is a consequence of the empirical character of this potential.
For that reason, Tran and Blaha have proposed the empirical alternative that prior to a band structure calculation with the mBJLDA potential, the lattice parameter 
is found from either a LDA or a GGA optimization procedure and the result introduced into the code to perform the band structure calculation of the semiconductor 
system. Such a procedure gives rise to quite improved results as compared to the previous version of the Wien2k code. It is known that the LDA underestimates as a 
rule, the lattice parameters and, on the contrary, GGA overestimates them. We have explored the possibility of using the averaged value as the lattice parameter, 
$a_{Avg}$, where $a_{Avg}=(a_{LDA}+a_{GGA})/2$. Here  $a_{LDA}(a_{GGA}$) is the lattice parameter obtained from an LDA (GGA) optimization procedure. When $a_{Avg}$ 
is used as input into the Wien2k code implemented with the mBJLDA potential, a better agreement of the band gap value with experiment is obtained as compared to 
the results with either $a_{LDA}$ or $a_{GGA}$. So this procedure turns out to give better results than the one recommended by Tran and Blaha and its extra 
computational cost is relatively low. A surprising result was, nevertheless, obtained when the experimental low temperature lattice parameter, $a_{LT}$, was 
introduced instead. Unexpected deviations of the band gap value from experiment as big as 48\% were obtained~\cite{20}. This is a disturbing result since the 
lattice parameters obtained from any optimization procedure are judged to be as good as the deviation from the experimental lattice parameter value is small, 
and so one expects to get the best result (the minimum deviation of the predicted band gap value from experiment) when the experimental lattice parameter is 
used. This is not the case. This fact throws doubts on the meaning of the optimization procedure altogether when the empirical mBJLDA code is employed. Nevertheless, 
we stress that the results obtained for the band gap value of semiconductors using the mBJLDA potential represents a relevant improvement at relatively low 
computational cost, a fact that we will emphasize below.

\section{The HSE method}

Hybrid functionals are a linear combination of Hartree-Fock (HF), LDA and GGA terms and were proposed initially with the aim of improving LDA and GGA in the calculation 
of the energy bands of molecules~\cite{22,23}. More recently, hybrid functionals were used as an effort to improve the old-standing problem of the band gap 
of semiconductors; they include the  Heyd-Scuseria-Ernzerhof (HSE) functional~\cite{13} proposed in 2003. It combines a screened short-range HF term and a screened 
short- and long-range functional proposed by Perdew, Burke and Ernzerhof (PBE)~\cite{4}. The screened terms in HSE result from splitting the Coulomb operator into 
short- and long-range terms in the following way
\begin{equation}
 \frac{1}{r}=\underbrace{\frac{erfc(\omega r)}{r}}_{SR}+\underbrace{\frac{erf(\omega r)}{r}}_{LR},
\end{equation}
where the complementary error function $erfc(\omega r)=1-erf(\omega r)$ and $\omega$ determines the range. The functional form of HSE is based on the hybrid
functional of Perdew, Burke and Ernzerhof (PBEh)~\cite{24} (also known in the literature as PBE1PBE and PBE0)~\cite{25,26}, as follows
\begin{equation}
 E_{xc}^{PBEh}=aE_{x} ^{HF} +(1-a)E_{x}^{PBE}+E_{x}^{PBE}+E_{c}^{PBE},
\end{equation}
The expression for the HSE exchange-correlation energy,  $E_{xc}^{HSE}$, is

\begin{eqnarray}
 E_{xc}^{HSE}=aE_{x} ^{HF,SR}(\omega) + (1-a)E_{x}^{\omega PBE,SR}(\omega)  \nonumber\\
   + E_{x}^{\omega PBE,LR}(\omega)+E_{c}^{PBE}, &
\end{eqnarray}

where $E_{x} ^{HF,SR}$ is the HF short-range functional (SR), $E_{x}^{\omega PBE,SR}$ and $E_{x}^{\omega PBE,LR}$ are the short-range (SR) and long-range (LR) 
components of the PBE functional. {\it a} is a mixing constant that is derived from perturbation theory~\cite{27}. In the literature, the functional HSE appears 
as HSE03 and HSE06. The difference is in the choice of the value of $\omega$. We will refer to the HSE03 simply as HSE, in this work. In 2005, Heyd et al.~\cite{14}
reported a study of the band gap and lattice parameters of semiconductor compounds using the HSE functional. We will comment on these results below.

Recently, Marques et al.~\cite{28} have proposed to relate the mixing constant {\it a} to dielectric properties of the solid. They took 
$a=1/\epsilon_{\infty}^{PBE}$. Their calculation using the hybrid functional PBE0 improves the predictions for the band gap value of the 21 semiconductors 
considered as compared to the original formulation. Furthermore, they used  $a\sim\bar{g}$, and $a\sim\bar{g}^4$ where $\bar{g}$ takes the form of the term 
in parenthesis in eq.(2). They introduced this form of the $a$ parameter into the hybrid functionals PBE0 and HSE06, respectively, and got an improved result. These 
proposals improve the performance of the hybrid functionals at no extra cost. We will comment further on these results below.

\section{The HISS potential}
Another successful potential to calculate the band structure of semiconductors is the middle-range hybrid exchange and correlation Henderdon-Izmaylov-Scuderia-Savia 
functional (HISS)~\cite{16,17}. It also uses the PBE potential but in a different way,
\begin{equation}
 E_{xc}^{HISS}=E_{xc}^{PBE}+c_{MR}(E_{x}^{MR-HF}-E_{x}^{MR-PBE})
\end{equation}
where the last two terms in parentheses are the middle-range (MR) exact exchange and middle-range PBE exchange energies, given by

\begin{eqnarray}
 \frac{1}{r}=\underbrace{\frac{erfc(\omega_{SR} r)}{r}}_{SR}+\underbrace{\frac{erf(\omega_{SR} r)}{r}}_{LR}  \nonumber\\
   +\underbrace{\frac{erf(\omega_{SR} r)-erf(\omega_{LR} r)}{r}}_{MR}, &
\end{eqnarray}

In 2012, Lucero et al.~\cite{29} reported their study of the band gap and lattice parameters of some semiconductor compounds using HISS, with 
$\omega_{SR}=0.84a_{0}^{-1}$, $\omega_{LR}=0.20a_{0}^{-1}$ y $c_{MR}=0.60$. These values were determined by fitting them to some atomization energies, barriers 
hights and values of the gap for some compounds.

\section{The GW approximation (GWA)}
As it is well known, the many-body Shr\"odinger equation contains the Coulomb interaction term which is a two-body potential and creates the difficulty to solve 
it for realistic systems. To address this problem, the Hartree-Fock Approximation (HFA) adds to the average Coulomb potential (the Hartree term) a non-local 
exchange potential which reflects the Pauli Exclusion principle. The energy gap of semiconductors predicted in this way turns out to be in most cases too large. 
This is due to the neglect of correlations or screening which are crucial in solids. To simulate the effect of correlations, Slater introduced the $X\alpha$ 
approximation which may be regarded as a precursor of the modern DFT. In DFT, the ground state energy can be proved to be a functional of the ground state density 
but the explicit form of the functional is not known. The minimization of the total energy functional with respect to the density gives the Kohn-Sham equations. 
The unknown exchange and correlation potential is approximated either by the local density approximation (LDA), the generalized gradient approximation (GGA) or 
the meta-GGA, among others, which describe metals well but fail to account for the band gap of semiconductors. The empirical mBJLDA potential is a response. 
An alternative way to deal with this problem is the GW approximation (GWA). It is derived from many-body perturbation theory~\cite{30}. The form of the self-energy 
in the GWA is the same as in the HFA but the Coulomb interaction is dynamically screened remedying the most serious deficiency of the HFA. The corresponding self-energy 
is therefore non-local and energy dependent. The Green function is obtained from a Dyson equation of the form $G=G_0+G_0\Sigma G$ where $G_0$  describes the direct 
propagation without the exchange and correlation interaction and $\Sigma$ contains all possible exchange and correlation interactions with the system that an 
electron can have in its propagation. 
The GWA may be regarded as a generalization of the HFA but with a dynamically screened Coulomb interaction. The non-local HFA is given by
\begin{equation}
 \Sigma^x({\bf r},{\bf r'})= \sum_{k_n}^{(occ)}\psi^*({\bf r})\psi({\bf r'})\nu({\bf r}-{\bf r'})
\end{equation}
Where $\nu({\bf r}-{\bf r'})$ is the bare Coulomb interaction. The GWA corresponds to replacing the bare Coulomb interaction $\nu$ by a screened interaction W. In the 
language of perturbation theory this corresponds to
\begin{equation}
\Sigma^{x}({\bf r},{\bf r'},\omega)=\frac{i}{2\pi}\int d\omega' G({\bf r},{\bf r'},\omega + \omega') W({\bf r},{\bf r'},\omega')
\end{equation}
For details see ref.~\cite{11}. We will analyze the results obtained for the semiconductor gap value using this approximation in what follows. 

In 2005, Rinke et al.~\cite{31} using the so called OEPx(cLDA)+GW approximation obtained a reasonable agreement with experiment when calculating the band gap of a 
certain number of semiconductors. In 2007, Shishkin et al.~\cite{32,33} using a self-consistent GWA (GWA + DFT),  and the self-consistent GW approximation with 
attractive electron-hole interaction, scGW(e-h) accounted quite well for the experimental band gap of several semiconductors. 

Now we proceed to analyze some of the different offers in the literature in what the calculation of the band gap of semiconductors is concerned and compare their 
results among themselves and with experiment.

\section{Analysis}
\begin{figure*}
\centering
     \includegraphics[width=1\textwidth]{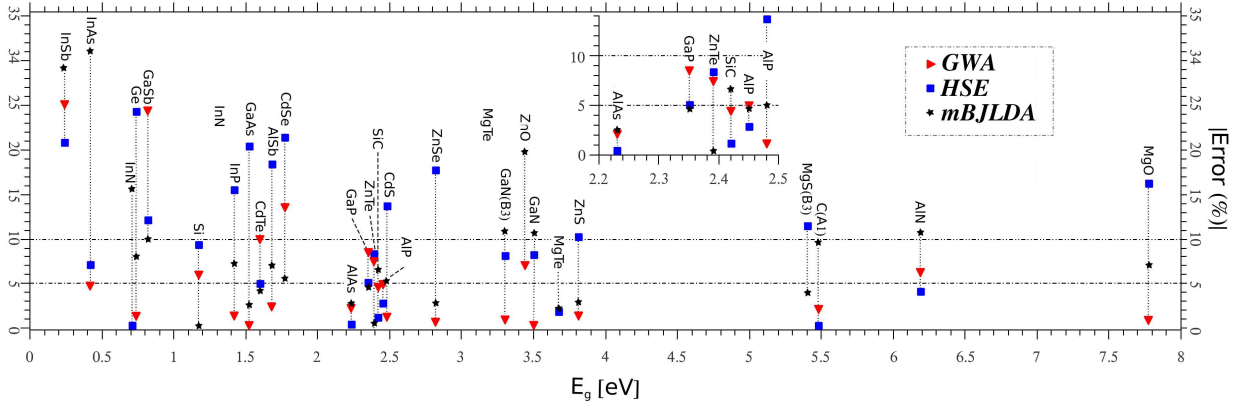} 
    \caption{\label{fig1}The horizontal axis is the absolute percent error ($ \left|Error(\%)\right|$). Each horizontal line is drawn at the low-temperature 
    experimental band gap value for each semiconductor. It compares, the three best results (GWA, mBJLDA, and HSE) according to our analysis. The inset 
    shows in more detail the semiconductors with a gap between 2.2 to 2.5 eV for clarity.(see Table \ref{tab:Tabla1} and text). }
\end{figure*}

In Table \ref{tab:Tabla1}, we present the results obtained for the band gap using different methods and approximations. The first two columns refer to our 
calculation. The resulting values using LDA and  mBJLDA~\cite{20} as in version 2011 of Wien2k code are presented. We have used as lattice parameter the average 
of the values obtained from an LDA and a GGA optimization which, as we found~\cite{20}, gives the best results for the gap as compared to 
experiment when the mBJLDA potential is used. Next we report the values obtained with the hybrid HISS and HSE06 functional~\cite{29}, HSE~\cite{13,14} and GWA. In 
the column denoted as GWA, we include the most precise predictions for the gap reported in the literature using either the self consistent GW 
(scGWA)~\cite{33,34,35,36,37,38,39,40} or the scGWA with attractive electron-hole interaction, scGW(e-h)~\cite{32}.
In Fig.~\ref{fig1}, the band gap value is given in the vertical axis. 
Each horizontal line is drawn at the experimental low temperature band gap value for each of the 27 semiconductors considered. 

\begin{table*}\footnotesize 
\centering
\caption{\label{tab:Tabla1}We compare the results for the gap ($E_g$), in eV, that we obtained with the Wien2k(LDA) code and with mBJLDA potential, with the hybrid 
functionals HISS, HSE06, HSE and with the GWA (see text). The crystal structure and percentage difference with respect to the experiment is shown in parenthesis.
The minus sign means that the calculation underestimates the experimental value. The experimental data are from refs.~\cite{41,42,43,44,45,46,47,48,49}.}
\begin{tabular*}{0.96\textwidth}{@{\extracolsep{\fill}}lccccccc}\hline\hline
        &  \multicolumn{5}{c}{Gap}\\\hline
Solid   & $E_g^{LDA}$    & $E_g^{mBJLDA}$ &  $E_g^{HISS}$   & $E_g^{HSE06}$   & $E_g^{HSE}$      &  $E_g^{GWA}$             & $E_{g}^{Expt.}$ \\\hline\hline
\multicolumn{8}{l}{ The experimental band gap at Low temperature}\\
\hline
C(A1)    & 4.16 (-24\%)   & 4.95 (-9.7\%)  & 6.11 (11.5\%)   & 5.42 (-1.1\%)   & 5.49 (0.2\%)    & 5.6$^\text{a}$ (2.2\%)    & 5.48  \\ 
Si(A1)   & 0.45 (-62\%)   & 1.17 (0.0\%)   & 1.45 (23.9\%)   & 1.22 (4.3\%)    & 1.28 (9.4\%)    & 1.24$^\dagger$ (6.0\%)    & 1.17   \\
Ge(A1)   & 0.00 (-100\%)  & 0.80 (8.1\%)   & 1.08 (45.9\%)   & 0.54 (-27.0\%)  & 0.56 (-24.3\%)  & 0.75$^\text{a}$ (1.4\%)   & 0.74   \\
MgO(B1)  & 5.00 (-36\%)   & 7.22 (-7.1\%)  & 7.87 (1.3\%)    & 6.40 (-17.6\%)  & 6.50(-16.3\%)   & 7.7$^\text{b}$ (-0.9\%)   & 7.77   \\
AlAs(B3) & 1.32 (-41\%)   & 2.17 (-2.7\%)  & 2.40 (7.6\%)    & 2.16 (-3.1\%)   & 2.24 (0.4\%)    & 2.18$^\text{c}$ (-2.2\%)  & 2.23   \\
SiC(B3)  & 1.30 (-46\%)   & 2.26 (-6.6\%)  & 2.74 (13.2\%)   & 2.32 (-4.1\%)   & 2.39 (-1.2\%)   & 2.53$^\dagger$ (4.5\%)    & 2.42   \\
AlP(B3)  & 1.43 (-42\%)   & 2.33 (-4.9\%)  & 2.71 (10.6\%)   & 2.44 (-0.4\%)   & 2.52 (2.9\%)    & 2.57$^\dagger$ (4.9\%)    & 2.45   \\
GaN(B3)  & 1.90 (-46\%)   & 2.94 (-10.9\%) & 4.05 (22.7\%)   & 2.97 (-10.0\%)  & 3.03 (-8.2\%)   & 3.27$^\dagger$ (-0.9\%)   & 3.30   \\
GaAs(B3) & 0.47 (-69\%)   & 1.56 (2.6\%)   & 1.86 (22.4\%)   & 1.18 (-22.4\%)  & 1.21 (-20.4\%)  & 1.52$^\ddagger$ (0.0\%)   & 1.52   \\
InP(B3)  & 0.45 (-68\%)   & 1.52 (7.0\%)   & 2.23 (57.0\%)   & 1.61 (13.4\%)   & 1.64 (15.5\%)   & 1.44$^\text{d}$ (1.4\%)   & 1.42   \\
AlSb(B3) & 1.14 (-32\%)   & 1.80 (7.1\%)   & 2.05 (22.0\%)   & 1.85 (10.1\%)   & 1.99 (18.5\%)   & 1.64$^\text{d}$ (-2.4\%)  & 1.68   \\
GaSb(B3) & 0.07 (-91\%)   & 0.90 (9.8\%)   & 1.31 (59.8\%)   & 0.70 (-14.6\%)  & 0.72 (-12.2\%)  & 0.62$^\text{d}$ (-24.4\%) & 0.82   \\
GaP(B3)  & 1.39 (-41\%)   & 2.24 (-4.7\%)  & 2.67 (13.6\%)   & 2.42 (3.0\%)    & 2.47 (5.1\%)    & 2.55$^\text{d}$ (8.5\%)   & 2.35   \\
InAs(B3) & 0.0 (-100\%)   & 0.55 (31.0\%)  & 0.93 (121.4\%)  & 0.36 (-14.3\%)  & 0.39 (-7.1\%)   & 0.40$^\text{d}$ (-4.8\%)  & 0.42   \\ 
InSb(B3) & 0.0 (-100\%)   & 0.31 (29.2\%)  & 0.80 (233.3\%)  & 0.28 (16.7\%)   & 0.29 (20.8\%)   & 0.18$^\text{d}$ (-25\%)   & 0.24   \\
CdS(B3)  & 0.93 (-63\%)   & 2.61 (5.2\%)   & 2.72 (9.7\%)    & 2.10 (-15.3\%)  & 2.14 (-13.7\%)  & 2.45$^\text{e}$ (-1.2\%)  & 2.48   \\
CdTe(B3) & 0.49 (-69\%)   & 1.67 (4.4\%)   & 2.00 (25.0\%)   & 1.49 (-6.9\%)   & 1.52 (-5.0\%)   & 1.76$^\text{f}$ (10.0\%)  & 1.60   \\
CdSe(B3) & 0.38 (-79\%)   & 1.87 (5.6\%)   & 1.90 (7.3\%)    & 1.36 (-23.2\%)  & 1.39 (-21.5\%)  & 2.01$^\text{f}$ (13.6\%)  & 1.77   \\
ZnS(B3)  & 2.08 (-45\%)   & 3.70 (-2.9\%)  & 4.12 (8.1\%)    & 3.37 (-11.5\%)  & 3.42 (-10.2\%)  & 3.86$^\ddagger$ (1.3\%)   & 3.81   \\
ZnSe(B3) & 1.19 (-58\%)   & 2.74 (-2.8\%)  & 2.93 (3.9\%)    & 2.27 (-19.5\%)  & 2.32 (-17.7\%)  & 2.84$^\text{f}$ (0.7\%)   & 2.82   \\
ZnTe(B3) & 1.20 (-50\%)   & 2.38 (-0.4\%)  & 2.77 (15.9\%)   & 2.16 (-9.6\%)   & 2.19 (-8.4\%)   & 2.57$^\text{f}$ (7.5\%)   & 2.39   \\  
MgS(B3)  & -              & 5.18 (-4.1\%)* & 5.17 (-4.3\%)   & 4.48 (-17.0\%)  & 4.78 (-11.5\%)  & -                         & 5.40   \\
MgTe(B3) & -              & 3.59 (-2.2\%)* & 3.91 (6.5 \%)   & 3.49 (-4.9\%)   & 3.74 (1.9\%)    & -                         & 3.67   \\
GaN(B4)  & 2.06 (-41\%)   & 3.13 (-10.6\%) & 4.23 (20.9\%)   & 3.14 (-10.3\%)  & 3.21 (-8.3\%)   & 3.5$^\text{g}$ (0.0\%)    & 3.50   \\
InN(B4)  & 0.03 (-96\%)   & 0.82 (15.5\%)  & 1.51 (112.7\%)  & 0.66 (-7.0\%)   & 0.71 (0.0\%)    & -                         & 0.71   \\
AlN(B4)  & 4.11 (-34\%)   & 5.53 (-10.7\%) & 6.62 (6.9\%)    & 5.50 (-11.2\%)  & 6.45 (4.2\%)    & 5.8$^\text{g}$ (-6.3\%)   & 6.19   \\
ZnO(B4)  & 0.76 (-78\%)   & 2.76 (-19.8\%) & -               & -               & -               & 3.2$^\dagger$ (-7.0\%)    & 3.44   \\
$\Delta(\%)$   & 60.2\%   & {\bf 8.4\%}    & {\bf 34.1\%}    & {\bf 11.5\%}    & {\bf 10.2\%}    & {\bf  5.7\%}              &  -     \\ 
\hline
\multicolumn{8}{l}{ The experimental band gap at room temperature}\\
\hline
BP(B3)   & 1.15 (-43\%)   & 1.83 (-8.5\%)  & 2.43 (21.5\%)   & 2.21 (10.5\%)    & 2.16 (8.0\%)    & -                                & 2.00   \\
BN(B3)   & 4.39 (-29\%)   & 5.85 (-5.6\%)  & 6.69 (7.90\%)   & 5.90 (-4.8\%)    & 5.99 (-3.4\%)   & 7.14 (15.2\%)                    & 6.20   \\
MgSe(B1) & 1.71 (-31\%)   & 2.89 (17.0\%)  & 3.05 (23.5\%)   & 2.58 (4.5\%)     & 2.62 (6.1\%)    & -                                & 2.47   \\
BaS(B1)  & 1.93 (-50\%)   & 3.31 (-14.7\%) & 3.61 (-7.0\%)   & 3.21 (-17.3\%)   & 3.28 (-15.5\%)  & 3.92 (1.0\%)$^\text{h}$          & 3.88   \\
BaSe(B1) & 1.74 (-51\%)   & 2.87 (-19.8\%) & 3.14 (-12.3\%)  & 2.80 (-21.8\%)   & 2.87 (-19.8\%   & -                                & 3.58   \\
BaTe(B1) & 1.37 (-56\%)   & 2.24 (-27.3\%) & 2.48 (-19.5\%)  & 2.22 (-27.9\%)   & 2.50 (-18.8\%)  & -                                & 3.08   \\
BAs(B3)  & 1.23 (-16\%)   & 1.72 (17.8\%)  & 2.14 (46.6\%)   & 1.89 (29.5\%)    & 1.92 (31.5\%)   & -                                & 1.46   \\
\hline\hline
\multicolumn{8}{l}{*with mBJ($a_{LDA}$). $^\dagger$scGW(h-e) in Ref.~\cite{32}. $^\ddagger$ scGW in Ref.~\cite{33}. $^\text{a}$Ref.~\cite{34}. $^\text{b}$Ref.~\cite{35}. $^\text{b}$Ref.~\cite{36}.}\\
\multicolumn{8}{l}{$^\text{d}$Ref.~\cite{37}. $^\text{e}$Ref.~\cite{38}. $^\text{f}$Ref.~\cite{39}. $^\text{g}$Ref.~\cite{40}. $^\text{h}$Ref.~\cite{50}. }\\
\multicolumn{8}{l}{$\Delta(\%)$ is the average of the absolute percent deviations.  }\\
\multicolumn{8}{l}{We calculate the percent deviation as follows $Error(\%)=(E_g^{Teo.}-E_g^{Expt})*100/E_g^{Expt}$.}\\
\end{tabular*}
\end{table*}

The horizontal axis represents the absolute percent error,($ \left|Error(\%)\right|$) calculated as shown at the bottom of Table I. The three data on each horizontal line 
correspond to the result obtained using the mBJLDA potential, the HSE method and the GWA. So, for a particular semiconductor, the graph compares directly the performance of 
each of the three best methods as found in this work. (see Table \ref{tab:Tabla1}). 

As it is very well known~\cite{7} and as it appears in Table \ref{tab:Tabla1}, the LDA does not reproduce the experimental values of the band gap of 
semiconductors. Furthermore, The Wien2k (LDA) code produces for MgS and MgTe a band structure which shows a direct band gap in contradiction with 
experiment~\cite{20}. 

The results of the predictions obtained with the GWA are the most accurate with an averaged error of 5.7\%. The empirical mBJLDA potential produces results with an 
averaged error of 8.4\%. Next, the errors obtained with the HSE potential result in an averaged error of 10.2\%, HSE06 (11.5\%) and HISS (34.1\%). The 
GWA, the  mBJLDA potential, and HSE functional do better than the ones reported by Marques et al.~\cite{28}. They get results with averaged errors 
16.5\%, 14.4\% y 10.4\% using the hybrid functional $PBE0_{\epsilon_\infty}$, $PBE0_{mix}$ and $HSE06_{mix}$, respectively. In this paper, the authors 
suggest the possibility  that the mixing parameter should be related to physical variables. 
The performance of the GWA is highly accurate, 88\% of the calculated results recorded here show less than 10\% error. This is to be compared to the one 
obtained when using mBJLDA (74\%), HSE (54\%), HSE (42\%), HISS (35\%). On the other hand, more than a 20\% deviation from experiment occurs when using 
mBJLDA in 7\% of the cases, and with the GWA in 8\% of the semiconductors studied,  which is to be compared with HSE (15\%), HISS (46\%). All together, 
the  best results are obtained when using the GWA; mBJLDA is next, but also HSE gives results with acceptable accuracy. A special case is the very-low-gap 
InSb. In this case none of the methods give less than a 20\% error although it would be more reasonable to judge these results from the absolute deviation 
in electron-volts rather than from the percent deviation (see Table I).
It is important for the overall picture to stress that the GWA and our calculations with the mBJLDA potential present deviations less than 5\% in 15 of the 27 
semiconductors considered. When the HSE potential is considered, 12 of the 27 present less than a 5\% deviation .

One more observation. In the previous analysis we took into account only low temperature band gap data. In Table I, we also present some calculations for 
which we did not found experimental reports at low temperature. Since the calculations are done at  0K, room-temperature measurements require extrapolation 
either using Varshni's law or a quadratic fit or any other suitable method which, in any case, generates an extra incertitude in the obtained 0K data. If 
we rather use the high temperature data, the HSE potentials give a better agreement with experiment.

Recently the mBJLDA potential has presented very good results in the study of complex systems. Is the case of the electronic, and magnetic 
features of the metal-insulator transition phase of VO2, which are well reproduced using the mBJ potential~\cite{A}. This result does not reproduces 
correctly using the hybrid functional HSE~\cite{B}.

\section{Conclusions}

 The accurate calculation of the band gap of semiconductors is a difficult task that has been the object of intense research with the result of important 
 progress during the last approximately ten years. As a continuation of our previos work (PRB) where we performed a detailed analysis of the performance 
 of the recently published modified Becke-Jonhson potential presented in this work our analysis of some different solutions and compare their results among them
 to pinpoint the actual accuracy of this empirical potential as componed to other methods.
 A group of 27 semiconductors (see Table I) for which we found low temperature  data on the band gap value were considered. 
 The results of the GWA, the Wien2k implemented with the mBJLDA potential, and codes using a hybrid functional, HSE, and HISS  were taken into consideration. 
 The results reported by Marques et al.~\cite{28} were found to be less accurate than the ones of the GWA, the mBJLDA potential and the HSE functional 
 (see text above for the precise definition used here). The GWA was found to give, all together the best results. The mBJLDA potential produces results 
 slightly less accurate and HSE comes next. The two first methods give quite good results (prediction better than 5\% for 15 of the 27 semiconductors 
 studied). In Fig.~\ref{fig1}, we compare the performance of the three best methods found in this analysis for each of the 27 semiconductors separately. 
 It is important to stress the empirical character of the mBJLDA potential because it prevents the consistent definition of the optimization procedure which  
 contrasts with the sound bases of the GWA. Even with the several theoretical non-properly solved issues, the mBJLDA potential gives rise to acceptable 
 predictions of the band gap value as compared to experiment.
 An extra factor to be taken into account is the computational cost. If one would seek for precision without taking this factor into account, the GWA is the method 
 to use. If one would prefer to sacrifice a little the precision obtained against the savings in computational cost, the mBJLDA potential seems the  
 appropriate method. In conclusion, we can typify the state of matters with respect to the calculation of the band gap of semiconductors as follows. A 
 quite precise method does exist, the GWA approximation. It's computational cost is higher. A relatively quicker code, the Wien2k implemented with the 
 mBJLDA potential, gives somehow less accurate results but quite acceptable at lower computational cost. Other methods do exist but are less accurate. 
 Very recently, the new approximation announced in the ref~\cite{51} was implemented in the Wien2k 12.1 code for public use. The new
 hybrid functional YS-PBE0 is ``equivalent'' to the HSE one, according to the authors. We will study this new functional in future work.

\section{Acknowledgments}
The authors acknowledge to the GENERAL COORDINATION OF INFORMATION AND COMMUNICATIONS TECHNOLOGIES (CGSTIC) at CINVESTAV for providing HPC resources on the Hybrid Cluster 
Supercomputer "Xiuhcoatl", that have contributed to the research results reported within this paper. J.A.C.M acknowledges the support of Conacyt M\'exico through a PhD scholarship.


\begin{thebibliography}{}

\bibitem{1} W. Khon and L. J. Sham,
{\em Phys. Rev.} {\bf 140}, A1133 (1965).

\bibitem{2} J. P. Perdew and Y. Wang,
{\em  Phys. Rev. B} {\bf 45}, 13244 (1992).

\bibitem{3} J. P. Perdew, S. Kurth, A. Zupan, P. Blaha,
{\em  Phys. Rev. Lett.} {\bf 82}, 2544 (1999).

\bibitem{4} J. P. Perdew, K. Burke and M. Ernzerhof,
{\em Phys. Rev. Lett.} {\bf 77 }, 3865 (1996).

\bibitem{5} J. P. Perdew, S. Kurth, A. Zupan, P. Blaha,
{\em Phys. Rev. Lett.} {\bf E78}, 1396 (1997).

\bibitem{6} J. Tao, J. P. Perdew, V.N. Staroverov and G. E. Scuseria, 
{\em Phys. Rev. Lett.} {\bf 91}, 146401 (2003).

\bibitem{7} J. P. Perdew, 
{\em Int. J. Quantum Chem.} {\bf 30}, 451 (1986).

\bibitem{8} G. A. Baraff, M. Schluter,
{\em Phys. Rev. B} {\bf 30}, 3460 (1984).

\bibitem{9} V. I. Anisimov, J. Zaanen, and O. K. Andersen, 
{\em Phys. Rev. B} {\bf 44}, 943 (1991).

\bibitem{10} L. Hedin, 
{\em Phys. Rev.} {\bf 139}, A796 (1965).
  
\bibitem{11} F. Aryasetiawany and O. Gunnarssonz,
{\em Rep. Prog. Phys.} {\bf 61}, 237 (1998).

\bibitem{12} W. G. Aulbur, L. Jonsson, and J. W. Wilkins,
{\em Solid State Phys.} {\bf 54}, 1 (2000).

\bibitem{13} J. Heyd, G. E. Scuseria, and M. Ernzerhof,
{\em  J. Chem. Phys.} {\bf 118}, 8207 (2003).

\bibitem{14} J. Heyd, J. E. Peralta, G.E. Scuseria, and R. L. Martin,
{\em J. Chem. Phys.} {\bf 123}, 174101 (2005).

\bibitem{15} J. Heyd, G. E. Scuseria, and M. Ernzerhof,
{\em J. Chem. Phys.} {\bf 124} 219906 (2006).

\bibitem{16} T. M. Henderson, A. F. Izmaylov, G. E. Scuseria and A. Savin,
{\em J. Chem. Phys.} {\bf 127} 221103 (2007).

\bibitem{17} T. M. Henderson, A. F. Izmaylov, G. E. Scuseria and A. Savin
{\em J. Theor. Comput. Chem.} {\bf4} 1254 (2008).

\bibitem{18} F. Tran and P. Blaha,
{\em Phys. Rev. Lett.} {\bf 102}, 226401 (2009)

\bibitem{19} P. Blaha, K. Schwars, G.K.H. Madsen, D. Kvasnicka, and J. Luitz,
{\em  WIEN2K:Full Potential-Linearized Augmented Plane waves and Local Orbital Programs for Calculating Crystal Properties}, edited by K. Schwars, 
Vienna University of Technology, Austria, (2001).
  
\bibitem{20} J. A. Camargo-Mart\'inez and R. Baquero,
{\em Phys. Rev. B} {\bf 86}, 195106 (2012).

\bibitem{21} A. D. Becke and M. R. Roussel,
{\em  Phys. Rev. A} {\bf 39}, 3761 (1989).

\bibitem{22} A. D. Becke,
{\em J. Chem. Phys.} {\bf 98}, 1372 (1993).

\bibitem{23} A. D. Becke,
{\em J. Chem. Phys.} {\bf 98}, 5648 (1993).

\bibitem{24} C. S. Wang and B. M. Klein,
{\em Phys. Rev. B} {\bf 24}, 3393 (1981).

\bibitem{25} M. Ernzerhof and G. E. Scuseria, 
{\em J. Chem. Phys.} {\bf 110}, 5029 (1999).

\bibitem{26} C. Adamo and V. Barone,
{\em J. Chem. Phys.} {\bf 110}, 6158 (1999).

\bibitem{27} J. P. Perdew, M. Ernzerhof, and K. Burke,
{\em J. Chem. Phys.} {\bf 105}, 9982 (1996).

\bibitem{28} M. A. L. Marques, J. Vidal, M. J. T. Oliveira, L. Reining, S. Botti,
{\em Phys. Rev. B} {\bf 83}, 035119 (2011).

\bibitem{29} M. J. Lucero, T. M. Henderson, and G. E. Scuseria,
{\em J. Phys.: Condens. Matter},  {\bf 24} 145504 (2003).

\bibitem{30} A. L. Fetter and J. D. Walecka,
{\em Quantum Theory of Many-Particle Systems.} Courier Dover Publications (2003).

\bibitem{31} P. Rinke et al.,
{\em New J. Phys.} {\bf 7}, 126 (2005).

\bibitem{32} M. Shishkin, M. Marsman, and G. Kresse,
{\em  Phys. Rev. Lett.} {\bf 99}, 246403 (2007).

\bibitem{33} M. Shishkin and G. Kresse,
{\em Phys. Rev. B}  {\bf 75}, 235102 (2007).

\bibitem{34} M.S. Hybertsen and S. G. Louie,
{\em Phys. Rev. B} {\bf 34}, 5390 (1986).

\bibitem{35} U. Sch\"onberger and F. Aryasetiawan,
{\em Phys. Rev. B} {\bf 52}, 8788 (1995).

\bibitem{36} R. W. Godby, M. Schl\"uter and L. J. Sham,
{\em Phys. Rev. B} {\bf 37}, 10159 (1988).

\bibitem{37} X. Zhu and S. G. Louie,1
{\em Phys. Rev. B} {\bf 43}, 14142 (1991).

\bibitem{38} M. Rohlfing, P. Kr\"uger and J. Pollmann,
{\em Phys. Rev. Lett.} {\bf 75}, 3489 (1995).

\bibitem{39} O. Zakharov, A. Rubio, X. Blase, M. L. Cohen and S. G. Louie,
{\em Phys. Rev. B} {\bf 50}, 10780 (1994).

\bibitem{40} A. Rubio, J. L. Corkill, M. L. Cohen, E. L. Shirley and S. G. Louie,
{\em Phys. Rev. B} {\bf 48} 11810 (1993).

\bibitem{41} S. Adachi,
{\em  Handbook on Physical Properties of Semiconductors, Vols. I, II and III.} Kluwer Academic Publishers (2004).

\bibitem{42} O. Madelung,
{\em  Data in Sciene and Technology, Semiconductors Group IV Elements and II-V Compounds}. Ed. Springer-Verlag (1991).

\bibitem{43} O. Madelung,
{\em Semiconductors: Data Handbook CD-ROM}. Ed. Springer-Verlag (2003).

\bibitem{44} D. Wolverson, D. M. Bird, C. Bradford, K. A. Prior, B. C. Cavenett,
{\em Phys. Rev. B}, {\bf 64}, 113203 (2001).

\bibitem{45} K. Watanabe, M. Th. Litz, M. Korn, W. Ossau, A. Waag et al.
{\em J. Appl. Phys. } {\bf 81}, 451 (1997).

\bibitem{46} M. Feneberg, J. Daubler, K. Thonke, R. Sauer, P. Schley, R. Goldhahn,
{\em Phys. Rev. B}, {\bf 77}, 245207 (2008).

\bibitem{47} H. Morkoc,
{\em Handbook of Nitride Semiconductors and Devices}, Ed. Wiley-VCH (2008).

\bibitem{48} D. M. Roessler and W. C. Walker,
{\em J. Phys. Chem. Solids}, {\bf 28}, 1507 (1967).

\bibitem{49} G. Ram\'irez-Flores, H. Navarro-Contreras, A. Lastras-Mart\'inez, R. C. Powell and J. E. Greene,
{\em Phys. Rev. B } {\bf 50}, 8433 (1994). 

\bibitem{50} L. Tie-Yu, C. De-Chan and H. Mei-Chun,
{\em Chinese Phys. Lett.} {\bf 23}, 943 (2005).

\bibitem{A} R. Grau-Crespo, H. Wang and U. Schwingenschlogl,
{\em Phys. Rev. B}, {\bf 86}, 081101(R) (2012).

\bibitem{B} Z. Zhu and U. Schwingenschlogl,
{\em Phys. Rev. B}, {\bf 86}, 075149 (2012).

\bibitem{51} F. Tran and P. Blaha,
{\em Phys. Rev. B}, {\bf 83}, 235118 (2011).

\end{thebibliography}
\end{document}